\documentclass{article}

\usepackage[pagenumbers]{wacv} 
\usepackage[utf8]{inputenc} 
\usepackage{graphicx}      
\usepackage{amsmath}       
\usepackage{float}
\usepackage[width=0.8\textwidth]{caption}

\usepackage[pagebackref,breaklinks,colorlinks]{hyperref}

\title{Audio Avatar Fingerprinting: An Approach for Authorized Use of Voice Cloning in the Era of Synthetic Audio}

\author{Candice R. Gerstner\\
Research Directorate and AI Security Center \\
National Security Agency\\
Fort George G. Meade MD, USA\\
{\tt\small crgerst@uwe.nsa.gov}
}

\begin{document}
\maketitle
\begin{abstract}
With the advancements in AI speech synthesis, it is easier than ever before to generate realistic audio in a target voice---one only needs a few seconds of reference audio from the target---quite literally putting words in the target person's mouth.
This imposes a new set of forensics-related challenges on speech-based authentication systems, videoconferencing, and audio-visual broadcasting platforms, where we want to detect synthetic speech. At the same time, leveraging AI speech synthesis can enhance the different modes of communication through features such as low-bandwidth communication and audio enhancements---leading to ever-increasing legitimate use-cases of synthetic audio. In this case, we want to verify if the synthesized voice is actually spoken by the user. This will require a mechanism to verify whether a given synthetic audio is driven by an authorized identity, or not. We term this task \textbf{audio avatar fingerprinting}.  As a step towards audio forensics in these new and emerging situations, we analyze and extend an off-the-shelf speaker verification model developed outside of forensics context for the task of fake speech detection and audio avatar fingerprinting, the first experimentation of its kind. Furthermore, we observe that no existing dataset allows for the novel task of verifying the authorized use of synthetic audio---a limitation which we address by introducing a new speech forensics dataset for this novel task. 
\end{abstract}    
\section{Introduction}
\label{sec:intro}
\begin{figure}[tb]
    \centering
    \includegraphics[width=0.5\textwidth]{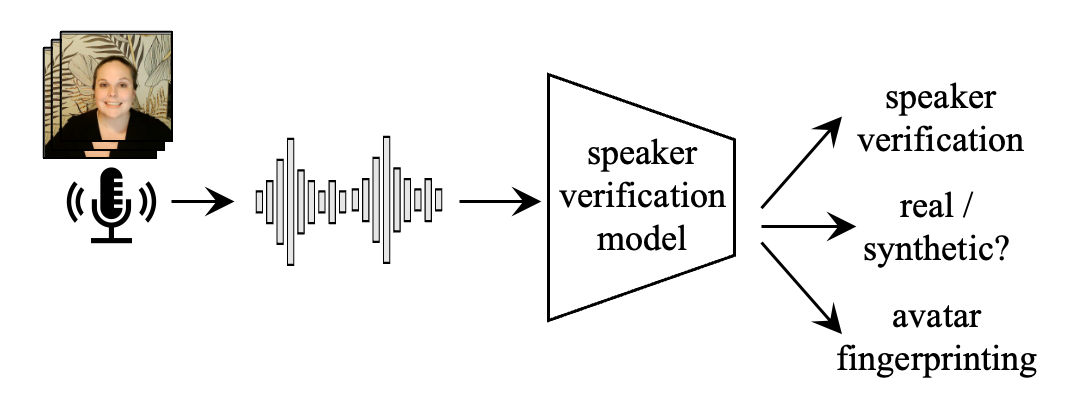}
    \caption{We explore how state-of-the-art off-the-shelf speaker verification models can be leveraged for audio forensics tasks: real vs. synthetic speech detection, and the newly introduced avatar fingerprinting task~\cite{prashnani2024avatar}, applied to audio. Note that the facial image is shown just to signify that the input the system is human speech: our approach is purely audio-based.}
    \label{fig:pipeline}
\end{figure}

The rapid advancement of generative AI has ushered in a new era of online human communication through a digital clone of ourselves that speak and look like us. State-of-the-art voice synthesis techniques can realistically replicate a person's voice using just a few seconds of a user's voice as reference. However, the growing accessibility of AI voice synthesis has also paved the way for its malicious use, enabling the creation of audio deepfakes for scams, political manipulation, and financial fraud ~\cite{NSA_FBI_CISA_2023}. In these scenarios, the central goal is to detect whether an audio sample is real or synthetic.    

On the other hand, advanced AI speech synthesis has potential to enable new real-time applications such as low-bandwidth communication, audio enhancements, or communication with avatars on social platforms~\cite{Veo3,heygen,MAXINE}. We are approaching a future where legitimate use cases of generated audio will become ubiquitous. This imposes a new set of audio forensics challenges to tackle, where the incoming audio may be known to be synthetic, but possibly not authorized. We term this task \textit{audio avatar fingerprinting}.

In this work, we investigate the performance of existing speaker verification models for the task of audio forensics (Fig.~\ref{fig:pipeline}). While speaker verification models are not originally designed for synthetic speech detection, our choice of speaker verification models is motivated by analogous tasks in video domains. In video forensics, identity recognition models have been successfully used by comparing unique motion features of that person to a fake video to determine if a video is real or fake~\cite{Agarwal_2019_CVPR_Workshops, Cozzolino_2021_ICCV}. Similarly, speaker verification models are trained to recognize a unique speech embedding of a user among real speakers in a form of identity recognition. Additionally, speaker verification models are typically trained using \textit{only} real speech of users without fake samples of any generators. This characteristic allows them to potentially generalize to arbitrary generators, making speaker verification models promising candidates for the task of audio forensics. A previous work~\cite{DeepfakePianese2022} has shown the effectiveness of speaker verification frameworks for real or synthetic speech detection using the audio component of an audio-video identity verification model~\cite{AVPOICozzolino2023}. In this work, we employ a classical, off-the-shelf speaker verification model developed outside of forensics context, and our studies go beyond spoof detection to its effectiveness in the emerging audio forensic task (i.e., audio avatar fingerprinting).

In this work, we analyze the speaker verification models in the following two tasks (see Fig.~\ref{fig:pipeline}). 
First we analyze how well they perform on the standard synthetic speech detection task: that is, how well does a state-of-the-art speaker verification model (TitaNet) discriminate between a real and a synthetic voice. Surprisingly, we find that the speaker embeddings in the vanilla TitaNet, which has never heard any fake audio during training and is trained for classical speaker verification tasks, can perform the fake audio detection task sufficiently well.

Second, we assess whether the learnt speaker representations are also able to detect whether a given synthetic audio is driven by an authorized identity, or not---a task that is adapted to audio from the recently-proposed ``avatar fingerprinting'' work in video forensics~\cite{prashnani2024avatar}.
In video-based avatar fingerprinting, a video avatar is analyzed to answer the question: ``is the talking-head video being driven by the person shown in the video or not?'' This is done by learning motion signatures that are independent of the facial appearance shown. In other words, the goal is to learn \textit{who} is driving the video regardless of what the avatar looks like. Equivalently, we want to explore how effective existing speaker representation models are for audio avatar fingerprinting: discerning whether the synthesized speech is driven by the authorized identity or not based solely on person-specific talking mannerisms, the audio counterpart of motion signatures. 
In other words, we seek to learn a talking-style signature of the driver---distinct from the acoustic appearance of the target voice---so that we can identify who is driving the voice regardless of how the voice sounds. 

To address this novel task of audio avatar fingerprinting, we introduce a new synthetic audio dataset since no existing datasets serve its training and validation requirements. To evaluate if the model correctly learns to extract unique talking signatures from the synthetic audio, we require synthetic voices for two cases: (i) self-reenactment, in which the target voice is driven by the same identity, and (ii) cross-reenactment, in which the target voice is driven by a different identity. Existing audio-spoof detection datasets primarily focus on cross-reenactment cases and no existing datasets contain both self- and cross-reenactment, which is crucial for this new task. 

Our preliminary results show that with some fine-tuning, the TitaNet embedding space can be adjusted to tackle the task of audio avatar fingerprinting, attributing who is driving the voice, regardless of how the voice sounds.

In summary, our contributions include:

\begin{itemize}

    \item we introduce a new audio forensics task of audio avatar fingerpriting and its dataset (Sec.~\ref{sec:dataset}) to support its training and validation.
    \item we analyze and extend the off-the-shelf speaker verification model for the task of fake speech detection (Sec.~\ref{sec:def-real-fake}) and audio avatar fingerprinting (Sec.~\ref{sec:def-af}) 
    \item our preliminary experiments show a promising avenue for both spoof detection and safe authorized use of AI voice synthesis technologies.
\end{itemize}

\section{Related Work}
\label{sec:related}
We provide an overview of the state-of-the-art in creating and detecting real-time deepfake audio and traditional speaker verification.

\subsection{Voice Synthesis}

	Synthesizing speech is not a new concept. In fact, people have been working on this technology since the 1700s~\cite{Brackhane2015}. This has traditionally been in the form of a text-to-speech (TTS) application. In 2016 with the advent of WaveNet~\cite{WaveNetOord2016}, these TTS systems quickly transformed to Deep Learning based approaches ~\cite{wang2023neural,chen2024vall-e, WaveGlowPrenger2018, FastSpeech2Ren2020} and have become more believable by the day. In addition to TTS, synthetic speech can also be created using Voice Cloning (VC) techniques. VC is the process in which speech is generated from existing audio to make it sound like an individual of choice. VC technologies offer the ability to extract the characteristics of a target voice in an attempt to make it sound more realistic. 

\subsection{Synthetic Voice Detection}

	Since the advent of synthetic speech technologies, people have been working on ways to detect them. The idea of spoof-detection came into the spotlight in big conferences like INTERSPEECH over ten years ago (2013) \cite{interspeech2013}. Ever since, people have been working on ways to detect synthetic speech in the context of automatic speaker verification (ASV). Competitions have also been built off of this concept, such as ASVSpoof (starting in 2015)~\cite{ASVSpoof2015}, SASV (starting in 2022)~\cite{Aljasem2021SASV}, and more recently SAFE (Synthetic Audio Forensics Evaluation) Challenge~\cite{dsri_safe_audio_forensics}. While some early work have relied on hand-crafted low-level features, such as constant Q cepstral coefficients (CQCC), Mel-frequency cepstral coefficients (MFCC), and jitter~\cite{jitter} etc, recent work leveraged deep neural network architectures including graph neural networks to capture spectro-temporal behaviors~\cite{Jung2021AASIST} or large pre-trained networks, such as wav2vec2~\cite{NEURIPS2020_92d1e1eb} and WavLM~\cite{wavlm}, to extract learned audio features~\cite{kang2024experimentalstudyenhancingvoice,MoPWav2vec,zhang2024audio,zhu2024slim} for spoof detection related tasks. Most of these methods require supervised training using real and fake audio datasets by known generators, and generalization to unseen generators has been a challenging task. These supervised methods will generally need to play a "cat-and-mouse" game indefinitely whenever new generators appear. Closest to our work is this previous work which presented the speaker verification framework for a fake audio detection task~\cite{DeepfakePianese2022} using a speech sub-module of an identify verification model jointly trained with both audio and videos~\cite{AVPOICozzolino2023}. On the other hand, our method employs an off-the-shelf, state-of-the-art classic speaker verification model and we study its performance for forensics tasks.

Other methods that have been developed over the years look to model the physical nature of the vocal track in order to determine the plausibility that the speech came from a human being as opposed to a computer program \cite{Blue2022Who, Puts2012, Smith2005}. These methods focus on orthogonal aspects to the traditional methods developed in the competitions described above.

\subsection{Speaker Verification}
Speaker verification is a biometric task that aims to confirm or deny a speaker's claimed identity based on their voice. The process typically involves two phases: \textit{enrollment} and \textit{verification}. During enrollment, a reference audio clip from a known user is processed to extract a unique ``voiceprint" or audio embedding, which serves as a biometric template for that individual. In the verification phase, a new audio clip from a test speaker (claiming a specific identity) is similarly processed to generate an embedding. This embedding is then compared against the enrolled reference embedding(s) for the claimed identity. The system is trained to differentiate between genuine users and imposters, primarily by classifying whether the incoming audio embedding belongs to the registered user or not. This training typically utilizes a large corpus of real audio clips from various speakers to learn robust and discriminative speaker characteristics.
Notable contributions include early work on Gaussian Mixture Models (GMMs) and i-vectors \cite{dehak2010front}, followed by the widespread adoption of deep learning approaches such as x-vectors \cite{snyder2018x}, and TitaNet \cite{Koluguri2022Titanet}. In this work, we leverage TitaNet that was trained with additive angular loss using depth-wise separable convolutional encoders as our speaker embedding model, benefiting from their proven effectiveness in generating robust and discriminative speaker embeddings. 

\subsection{Avatar Fingerprinting}

The concept of avatar fingerprinting is presented by previous work in the context of verifying the authorized use of portrait video avatars~\cite{prashnani2024avatar}. In this previous paper, the researchers developed a method that would help distinguish who the driver behind a synthetic talking-head video was based on person-specific talking mannerisms (e.g., person's facial expressions). In this task, they assume that the input media is \textit{known to be synthetic} and aim to verify if the authorized user is driving the portrait of themselves or not. They were able to learn an embedding space in which the motion signatures of one identity were grouped together, and pushed away from those of other identities, regardless of the appearance in the synthetic video. To our knowledge, this concept was the first of its kind in the video domain and we are extending that idea in the audio domain in this paper.
\section{Method}
\begin{figure}[tb]
    \centering
    \includegraphics[width=0.5\textwidth]{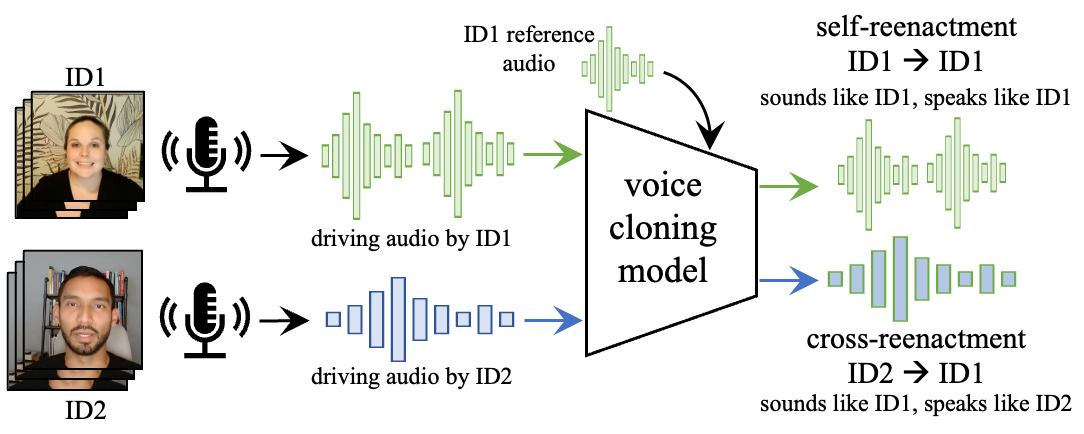}
    \caption{Voice cloning generators require a short reference sample of a target voice (this is what the final output will sound like), and a driving audio speech (this is what the target voice will speak). If a person is using their own voice as driving and reference (target) audio, then the resultant generated audio is termed a ``self-reenactment'': that is, the sound of the generated voice matches that of the driving audio. On the other hand, if one uses a different reference voice than that of the driving speaker, then the resultant generated audio is termed a ``cross-reenactment'': that is, the sound of the generated voice is different than that of the driving speaker, and it seems like the target / reference voice is speaking the words of the driving speaker.
    Here, we show a case of self- and cross-reenactment of ID1.}
    \label{fig:definition}
\end{figure}

\begin{figure}[b]
    \centering
    \includegraphics[width=0.5\textwidth]{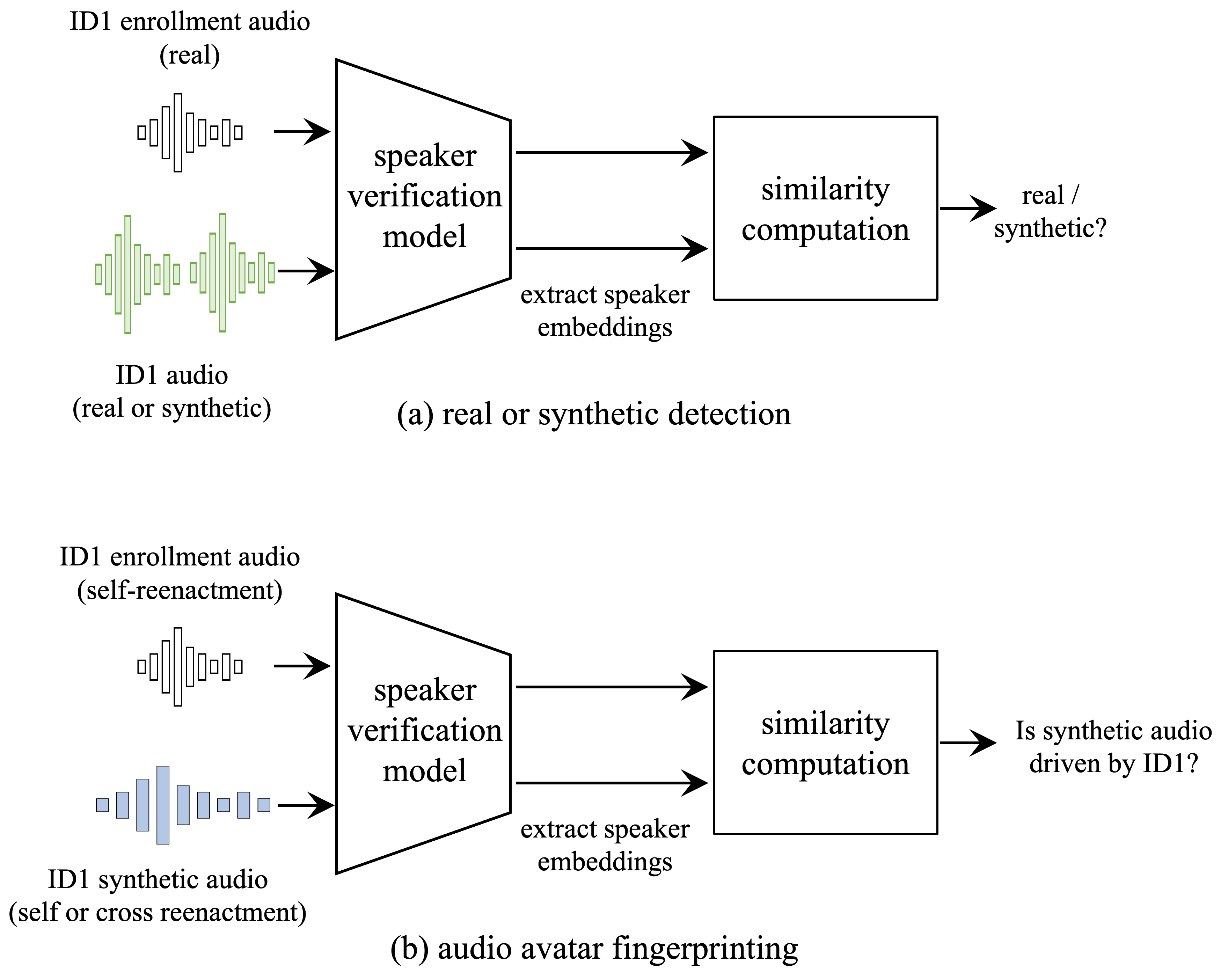}
     \caption{We explore the performance of off-the-shelf speaker verification models on two tasks. First (a), we want to evaluate whether the learnt speaker representation can be leveraged for detecting real and synthetic audio of a speaker (regardless of whether the synthetic audio is self- or cross-reenacted). See Sec.~\ref{sec:def-real-fake}. Second (b), we want to explore whether the speaker representations can be used for audio avatar fingerprinting: that is, identifying if the speaker driving a synthetic audio is the authorized identity or not (Sec.~\ref{sec:def-af}).}  
  \label{fig:definition_task}
\end{figure}

\subsection{Terminology}
\label{sec:terminology}
We first describe the definition of self- and cross-reenactment in the context of fake speech detection and audio avatar fingerprinting (see Fig.~\ref{fig:definition}). 
In the synthetic audio generation process, a ``target audio identity'' represents the sound of the generated voice.
That is, the final generated audio has the acoustic properties of the person whose voice is used as the target.
Typically, the voice cloning generator is provided with a sample audio of this target / reference voice.
Then, a driving audio speech is used to ``drive'' the target voice: the content (words) spoken and the talking mannerism is determined by this driving speech. 
When a person is using their own voice as driving and reference (target) audio, we call the resultant synthetic audio a ``self-reenactment''.
In contrast, if one uses a different reference voice than that of the driving speaker, then the resultant generated audio is termed a ``cross-reenactment''. Such an audio gives the impression that the target / reference voice is speaking the words of the different driving speaker. 
We use the notation $\text{ID}_{i}\rightarrow \text{ID}_{j}$ to indicate when synthetic audio that sounds like the voice of $j$ is driven by an identity $i$ throughout the paper. Please refer to Fig.~\ref{fig:definition} for illustration. 

Next, we discuss our approach for spoof detection and the audio avatar fingerprinting task (Fig.~\ref{fig:definition_task}).
	
\subsection{Real or Synthetic Speech Detection}
\label{sec:def-real-fake}
In this scenario, we consider synthetic to be audio that was generated using either text-to-speech (TTS) or voice cloning (VC). We do not consider the edge case in which there are legitimate uses for synthetic speech, such as an assistive technology for someone that is unable to speak. 

We assume enrollment audio is available for a speaker verification network and therefore will only discuss that situation here (see Fig.~\ref{fig:definition_task}(a)). Given one or more samples of verified real enrollment audio, an out-of-the-box speaker verification model can be used to compute an average speaker embedding for the enrollee. For the speaker verification model, we use TitaNet, which is a state-of-the-art speaker verification model trained on large-scale real speech data. 
Following Fig.~\ref{fig:definition_task}, we extract a speaker embedding for an input audio clip, which may be real or synthetic audio of the user. Given the pair of speaker embeddings of the verified enrollment audio and audio under question, we compare them using cosine similarity to determine if the incoming audio belongs to the user or not, thereby classifying non-matching audio as synthetic. We discuss the results of the task in Sec.~\ref{sec:result-real-fake}.

\subsection{Audio Avatar Fingerprinting}
\label{sec:def-af}
	 
\begin{figure}[b]
    \centering
    \includegraphics[width=0.5\textwidth]{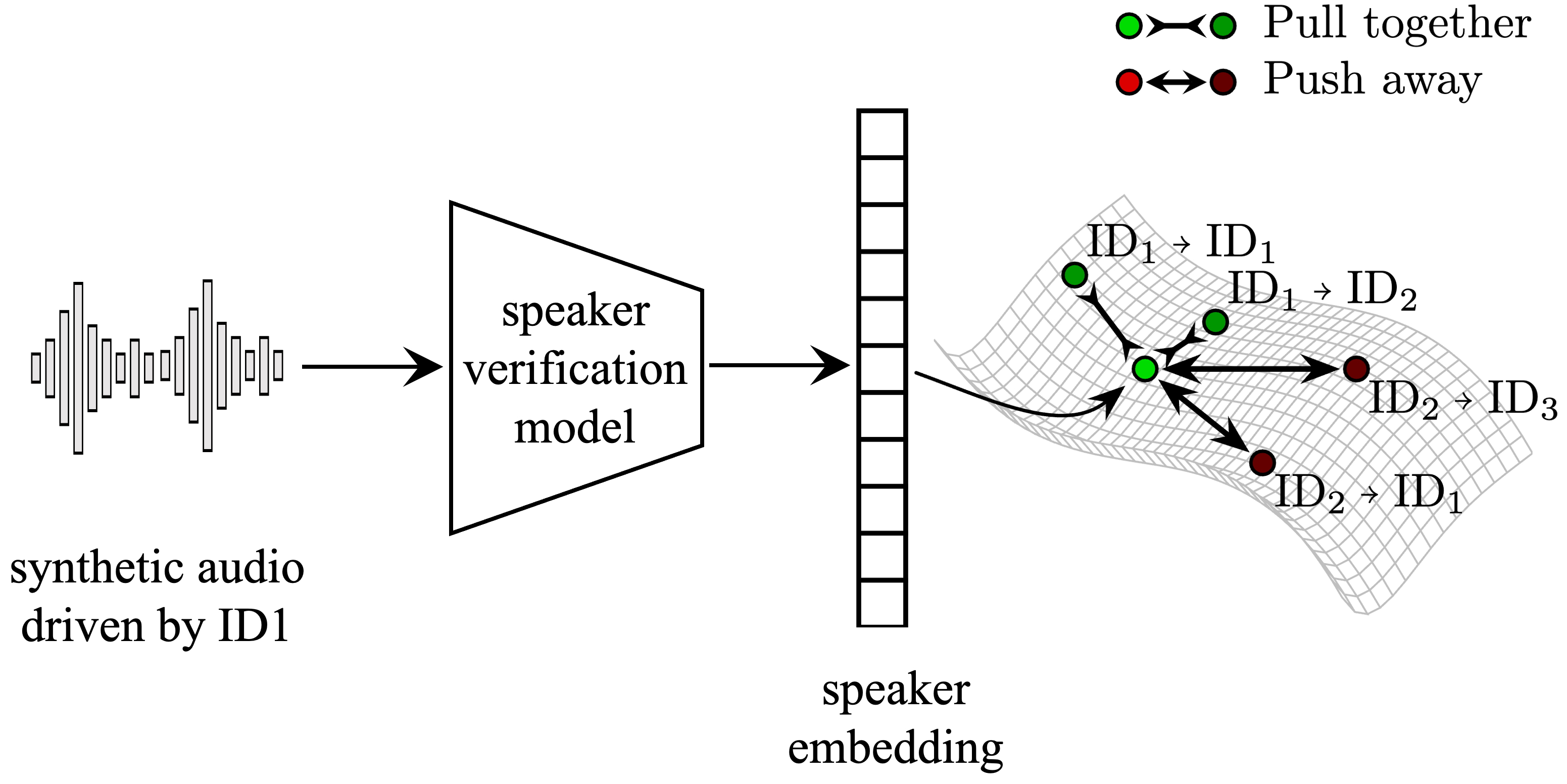}
    \caption{We use a speaker verification model (TitaNet) to extract a speaker embedding for a given synthetic audio. To learn a speaker embedding that encodes the speaking style of the driving identity, we use an additive angular margin loss that clusters all the audio driven by ID1 to be closer and away from those driven by others.}  
    \label{fig:contrastive}
\end{figure}
In this task, our goal is to determine if the input synthetic speech is driven by the authorized identity, analogous to the original avatar fingerprinting task in the visual domain~\cite{prashnani2024avatar}. Recall that in this context, a self-reenactment refers to a case in which the identity providing the sound of the voice matches the identity driving it, while a cross-reenactment refers to a case in which the sound of the generated voice is different from that of the driving speaker (see Sec.~\ref{sec:terminology}). With this in mind, we can formally state our task is to determine if the synthesized audio is a self- or cross-reenactment. 
Fig.~\ref{fig:definition_task}(b) shows our pipeline for the audio avatar fingerprinting task. The overall pipeline is similar to the task of real or synthetic detection in the previous section, but now the enrollment audio is the self-reenactment of $\text{ID}_{1}$, and the incoming audio is synthetic audio of $\text{ID}_{1}$, which may be self- or cross-reenactment. In this setting, since the input audio to the model is known to be synthetic and always sounds like the voice of the target identity, we cannot rely on the acoustic property of the synthesized audio to classify the driving identity. Instead, an ideal approach should rely on the speaking style (e.g., talking mannerism) of the synthetic audio to verify the driving identity. We wish to use TitaNet's speaker embedding to discriminate talking styles of different driving identities, but as we show in our preliminary analyses in Sec.~\ref{sec:result-af}, out-of-the-box TitaNet embeddings favor how the voice sounds rather than the talking mannerism, and are not suitable as-is for fingerprinting the driver. For TitaNet to cluster audio based on a driver identity as opposed to a target identity, we finetune TitaNet based on an angular softmax loss that pulls all the audio driven by $\text{ID}_{i}$ to be closer and pushes away those driven by others. See Fig.~\ref{fig:contrastive} for the illustration.  %

We discuss the results of the task in Sec.~\ref{sec:result-af}.
\section{The NVFAIR Audio Dataset}
\label{sec:dataset}

In this section we will discuss the production of a dataset that was used to evaluate our tasks on both spoof detection and audio avatar fingerprinting. Recall, to evaluate our novel task on audio avatar fingerprinting, we need a dataset with both self- and cross-reenactment. 
We use the NVFAIR Dataset~\cite{prashnani2024avatar} as a basis and augment it to generate the audio data necessary for all our tasks. The original audio was derived from video conferencing recordings of 46 subjects of diverse genders, age groups, and ethnicities. 
These recordings had two distinct recording strategies: a free-form stage where the subjects were given only general guidance on the topics, and a more controlled scripted stage. 
The audio was extracted from these videos and used as input to a state-of-the-art VC technique~\cite{Voicefont} in which synthesized speech was developed for each audio sample for each of the 46 target identities. 
That is, each identity drove themselves (self-reenactment) \textit{and} all other identities (cross-reenactment). 

In summary, with this generation process, we create a dataset of synthetic VC audios, with 4 hours 45 minutes of self-reenacted audio across 1899 unique audio files, and 90 hours 12 minutes of cross-reenacted audio across 74331 unique audio files.

\section{Results}

We evaluate the efficacy of our technique on multiple in-the-wild datasets. 

\subsection{Real or Synthetic Speech Detection}
\label{sec:result-real-fake}
\paragraph{Dataset. }
We tested TitaNet against the newly introduced NVFAIR audio dataset (Sec.~\ref{sec:dataset}) and another recent benchmark dataset which we shall refer to as InTheWild \cite{in_the_wild_audio_deepfake}. This InTheWild dataset consists of audio from 58 celebrities and politicians and has 20.8 hours of bona-fide and 17.2 hours of spoofed audio for each. InTheWild contains real-world deepfake clips collected from online sources (219 source videos for the fake class). The specific speech generation systems used to create these deepfakes are not labeled / are unknown, so InTheWild serves as an open-set, real-world benchmark with unknown (potentially heterogeneous) generation methods. In these studies, we 
used real samples for enrollment data and evaluated TitaNet based on this reference information.

\paragraph{Evaluations. }
Tab.~\ref{tab:detection_table} shows comparisons to recent speaker verification-related methods in the context of spoof detection. Our average AUC across these tests for determining real or fake was 91\% across all speakers in the dataset. These were unexpected results given that TitaNet was trained for a classical speaker verification task, not trained for any forensics tasks. 
\begin{table}[h!]
  \centering
  \begin{tabular}{@{}lc@{}}
    \toprule
    Method & AUC \\
    \midrule
    ClovaAI~\cite{Heo2020clova} & 0.74 \\
    H/ASP~\cite{Heo2020clova} & 0.85 \\
    ECAPA-TDNN~\cite{desplanques2020ecapa} & 0.79 \\
    POI-Forensics~\cite{AVPOICozzolino2023} & \textbf{0.91} \\
    Ours & \textbf{0.91}\\
    \bottomrule
  \end{tabular}
  \caption{Comparisons of synthetic speech detection task on InTheWild datasets \cite{in_the_wild_audio_deepfake} with other speaker verification methods. The numbers are cited from ~\cite{DeepfakePianese2022}. }
  \label{tab:detection_table}
\end{table}

\begin{figure}[t]
    \centering
    \includegraphics[width=0.5\textwidth]{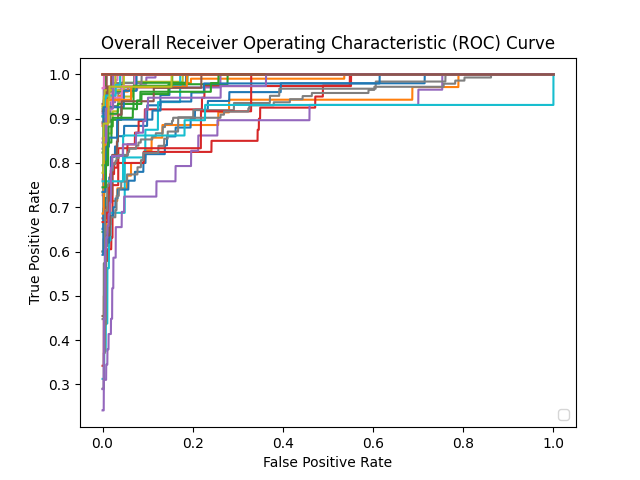}
    \caption{Real vs fake test with \textit{vanilla} TitaNet on NVFAIR}
    \label{fig:realfake}
\end{figure}	

Our next experiments went to evaluating an even more up-to-date dataset based on state-of-the-art VC technology~\cite{Voicefont}. Our tests with the NVFAIR dataset provided us with substantial data to continue to test our hypothesis that TitaNet could be used effectively to determine if an audio clip was real or fake. For this test, we used an average of 2 minutes of enrollment data for each speaker as a reference embedding and used TitaNet to compare embeddings of both real and fake data of that speaker, which again were generated with a state-of-the-art VC technology. We achieved an average AUC of 0.98 across all 46 speakers in this test. See Fig.~\ref{fig:realfake}. 

\paragraph{Discussions on robustness. }Both InTheWild and newly introduced NVFAIR audio dataset represent challenging real-world scenarios with varying numbers of identities, varieties of audio environments and compressions including fake audio generated by multiple (unknown) generators. Our results demonstrate that an out-of-the-box speaker verification model (i.e., TitaNet), trained on a large-amount of in-the-wild real audio, can produce an audio embedding sufficiently reliable and robust to discriminate fake audio from the real ones.

\subsection{Audio Avatar Fingerprinting}
\label{sec:result-af}

\paragraph{Dataset. } 
We used the newly constructed NVFAIR audio dataset to evaluate this task. 

In the case of the self-reenactment category for avatar fingerprinting, the driver identity was the same as the target identity. For example, $\text{ID}_{1}$ was used as a driver to generate cloned audio of themselves. This was done for each of the 46 identities in the dataset. 
In the case of the cross-reenactment category for avatar fingerprinting, the driver identity remained the same and the target identity was varied. For example, $\text{ID}_{1}$ was used as a driver to generate cloned audio of the 45 other identities in the dataset.
We split the data so that 40 identities were used for finetuning and the remaining 6 identities for testing. 
Of the 40 identities, we removed all instances of cross re-enacted data for the remaining 6 identities in the training data. That is, the test-set identities did not overlap with the training set, \textit{both} in terms of the driving identities and target identities. We then held out 5\% of the training data as a validation set. The data was labeled to uniquely identify each audio by its driving identity.
This allowed us to structure the embedding space such that the audios clustered based on the driving identity audio, regardless of the target identity. 

\begin{figure}[t]
    \centering
    \includegraphics[width=0.5\textwidth]{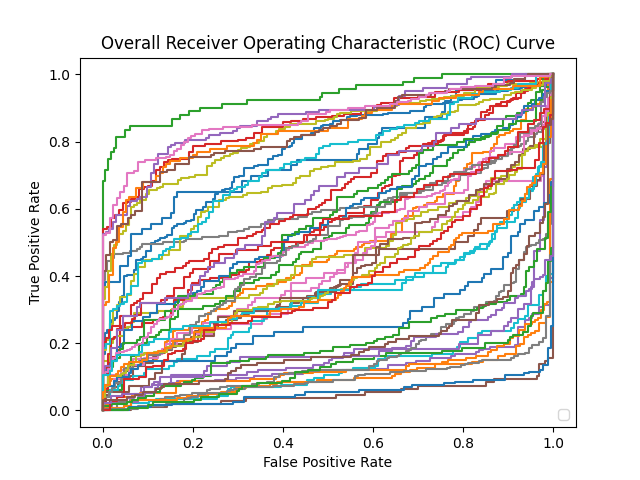}
    \caption{A preliminary analysis of a \textit{vanilla} TitaNet for an avatar fingerprinting task.}
    \label{fig:af_vanilla1}
\end{figure}
\begin{figure}[t]
    \centering
    \includegraphics[width=0.5\textwidth]{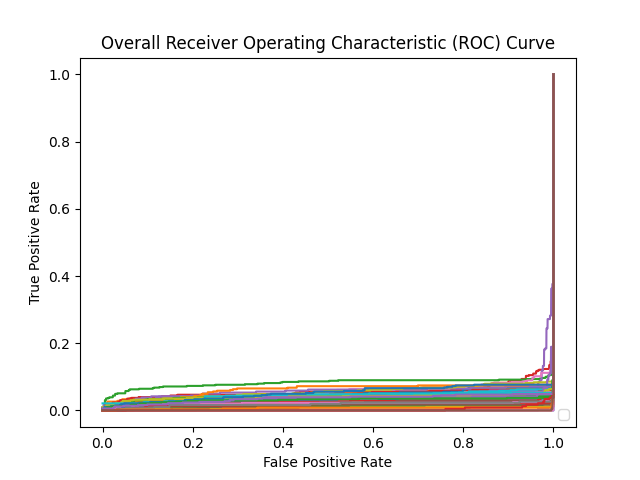}
    \caption{A preliminary analysis of a \textit{vanilla} TitaNet for an avatar fingerprinting task using only self-reenactment as authorized enrollment data}
    \label{fig:af_vanilla2}
\end{figure}

\paragraph{Preliminary analyses of vanilla TitaNet embeddings.}
Since TitaNet was performing better than expected on the tasks such as distinguishing real vs fake, we decided to try extending to the case of avatar fingerprinting. In this case, all audio seen by the model was synthetic and the task was to classify the audio based on the driver as opposed to what the audio sounds like (the target voice). As a preliminary manner, we conducted two analyses using a vanilla TitaNet to understand its out-of-the-box behaviors for this novel task (see Figs.~\ref{fig:af_vanilla1} and \ref{fig:af_vanilla2}). 

Our first analysis took a random sample of audio (approximately two minutes) driven by a given identity regardless of target identity for enrollment data. 

For example, if $\text{ID}_{1}$ is the authorized enrolled user, then we chose samples of synthesized audio where $\text{ID}_{1}$ was driving the synthesized voice.

Based on the average embedding generated by TitaNet for the enrollment data, which included both self and cross-reenacted data of the given identity, 
the embeddings of the test data were then compared to the average enrollment embedding using cosine similarity and a ROC curve was generated for each of the 46 identities. The result is shown in Fig.~\ref{fig:af_vanilla1}. This approach provided some results that are worthy of further exploration. In particular, exploration for the identities in which the embeddings consistently provided the opposite result (those that are in the lower right section of the ROC curve). 

For some failed individuals (lower right), they have too distinctive voice style (as in "appearance" in the video) or did not have any distinctive mannerism, which confused the avatar fingerprinting model while other successful individuals (top left), they have distinctive talking mannerism and the speaker verification model was able to capture it. 

In our second analysis to this problem, we limited the enrollment data to authorized \textit{self-reenacted} data, where the synthesized voice is that of the authorized driver to further understand the model's behaviors. Based on this new embedding, we repeated the same process as the first experiment to determine if TitaNet can correctly attribute the driver in the audio regardless of how it sounds. 

In this case, based on the limited variations in enrollment data, we assume that TitaNet will perform poorly in this avatar fingerprinting task, as the model will latch onto how the voice sounds rather than the mannerisms of the driving speaker, leading to incorrect attribution of the driver. The results are shown in Fig.~\ref{fig:af_vanilla2}. 

Our intuition was consistent with the results of these experiments. We found that audio that was self-driven was classified correctly (ie - that of the identity), but audio that was cross-reenacted (driven by the identity but sounding like a different target) was misclassified by our definition. Also, audio driven by another identity to sound like the enrolled user was also misclassified as authorized. This is again expected because we only enrolled self-reenacted synthetic audio of the individual, but were trying to see if the model would pick up any characteristics of the driver. 
\paragraph{Analysis of fine-tuned TitaNet embeddings.}

\begin{figure*}
  \centering
  \begin{subfigure}{0.5\textwidth}
    \includegraphics[width=\columnwidth]{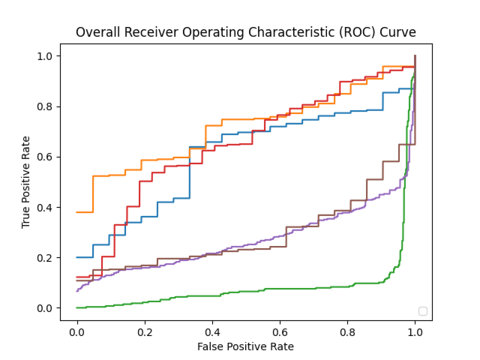}
    \caption{Before finetuning}
    \label{fig:af-before}
  \end{subfigure}
  \hfill
  \begin{subfigure}{0.5\textwidth}
    \includegraphics[width=\columnwidth]{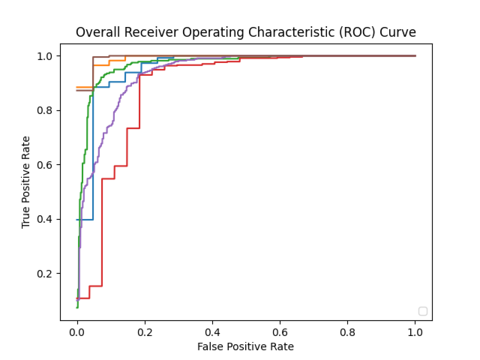}
    \caption{After finetuning}
    \label{fig:af-after}
  \end{subfigure}
  \caption{ROC curve of 6 held-out unseen identities from NVFAIR dataset using out-of-the-box TitaNet (a) and  fine-tuned TitaNet (b)}
  \label{fig:af}
\end{figure*}
The results of these tests led us to fine-tune TitaNet. Since the embeddings of out-of-the-box TitaNet seemed to be identifying relevant characteristics for some identities (in the first experiment), it was our hypothesis that with a little guidance, the embedding space could adjust to the relevant features for avatar fingerprinting. Using the method described in Sec.~\ref{sec:def-af}, we now show results for the held-out unseen 6 identities in the NVFAIR dataset. For reference, the first ROC curve provided in Fig.~\ref{fig:af-before} is the out-of-the-box TitaNet results for these identities. In Fig.~\ref{fig:af-after}, after fine-tuning, we were able to show promising results that the model is able to identify underlying audio properties of the driver. Our preliminary results show that fine-tuning TitaNet for given authorized identities is a promising solution for the problem of avatar fingerprinting in which synthetic audio is the assumed input.

\paragraph{Implementation details. } We finetune TitaNet for 10 epochs, using an Angular Softmax Loss and an Adam Optimizer, following all hyperparameters in the original paper. %

\section{Limitations and future work}
As a speaker verification method, our model assumes an availability of enrollment data. As with the classic speaker enrollment methods, even in a vanilla use case, it may be difficult to capture reliable enrollment data as the person's speaking style may vary based on their emotion or environments. While our preliminary work has successfully demonstrated the audio avatar fingerprinting task at small scale, it would be interesting in the future to train this model using a large number of subjects. In this work, we use TitaNet as a state-of-the-art speaker verification model. It would be fruitful future work to repeat the experiments we presented using a different speaker verification model to check if this concept generalizes to other models.

\section{Conclusion}

We will soon be in a world where everything in the digital world is synthesized in some way by AI. Therefore, the ability to verify the driver behind the avatar and transparency of the creation is extremely important in gaining trust. 

In this work, we have shown preliminary results on leveraging and extending an off-the-shelf speaker verification model for the tasks of real or synthetic speech detection and verification of authorized uses of AI voice synthesis. 
While the speaker verification model is not trained in the context of forensics, we show that they have a promising capability to generalize to spoof detection of arbitrary generators in the wild. We also show that with some slight tuning of the model, it can be adapted for the task of avatar fingerprinting. We hope the methods we have presented in this paper are a step toward enabling the advancement of online communication tools, such as video conferencing, using AI voice synthesis for audio enhancement and speech translation, while at the same time, ensuring the safe authorized use of AI voice synthesis methods. We believe that such methods, as the one described in this paper, will continue to grow in importance given the rapid evolution of generative AI technology and therefore we challenge the forensics community to develop additional methods in this novel area. 

\section*{Acknowledgements} The author would like to thank Nithin Koluguri, Koki Nagano and Ekta Prashnani at NVIDIA Research for many useful discussions on this work. This research was supported in part by NSA's Senior Technical Development Program (STDP).

\bibliographystyle{plain}
\bibliography{main}

\end{document}